\documentclass{aa}
\usepackage{graphicx}
\newcommand{\kms}{km\,s$^{-1}$}
\newcommand{\Rsun}{R$_\odot$}
\newcommand{\Msun}{M$_\odot$}

\begin{document}
 
%\title{A study of the young quadruple system AO\,Vel with a ZAMS eclipsing BpSi primary. 
%\title{The young quadruple system AO\,Velorum with a ZAMS eclipsing BpSi primary.
\title{AO Velorum: a young quadruple system with a ZAMS eclipsing BpSi primary.
\thanks{Based on observations obtained at the European Southern Observatory, 
La Silla, Chile (ESO programmes Nos.~072.D-0235, 65.L-0316), and at the 
Complejo Astron\'omico El Leoncito (CASLEO), Argentina.}
}

\author{
J.~F. Gonz\'alez\inst{1}
\and S. Hubrig\inst{2}
\and N. Nesvacil\inst{2,3}
\and P. North\inst{4}
}
\institute {Complejo Astron\'omico El Leoncito, Casilla 467, 5400 San Juan, Argentina
\and
European Southern Observatory, Casilla 19001, Santiago 19, Chile
%%%2nd official afiliation of Nicole:
\and
Department of Astronomy, University of Vienna, Tuerkenschanzstrasse 17, 1180 Vienna, Austria
\and 
Ecole Polytechnique F\'ed\'erale de Lausanne (EPFL), Laboratoire d'Astrophysique, 
Observatoire, CH-1290 Sauverny, Switzerland
}
\titlerunning{Quadruple system AO\,Vel}
\authorrunning{J. F. Gonz\'alez et~al.}
\abstract{ 

Using recent spectroscopic observations, we show that the triple system
AO\,Vel with an eclipsing BpSi primary is in fact a remarkable quadruple system 
formed by two double-lined spectroscopic binaries with components close to the ZAMS.
All available data have been reanalyzed to derive proper 
orbital parameters for both binary systems and to calculate
absolute parameters of the eclipsing system. For the first time, direct determination
of the radius and the mass have been obtained for a BpSi star.
The study of the physical parameters of this unique system is especially important since it can 
be used as a test of evolutionary models for very young stars of intermediate mass.

}

\maketitle

\section{Introduction}

Upper main sequence stars with chemical peculiarities and large-scale 
organized magnetic fields constitute about 10\% of A and B-type stars. 
To understand the physics of such stars, it is most important to identify
the origin of their magnetic field. Two main hypotheses have been proposed:
either these stars have retained the
primordial field present at their formation, the ohmic decay being
longer than the main sequence stellar lifetime (fossil field theory),
or the field is generated
and maintained by a contemporary dynamo. The first hypothesis implies that
strongly magnetic stars must be found on the ZAMS (as well as over the whole
width of the main sequence), while this is not necessarily true according
to the second hypothesis. Hubrig et al. (\cite{hu00}) suggested that magnetic
stars tend to accumulate in the middle of the main sequence strip if their mass
is below 3\,M$_\odot$ (which would challenge the fossil field theory), while
more massive magnetic stars seem more evenly distributed over the whole width of
the main sequence, including near the ZAMS (Hubrig et~al.\ \cite{hu05}).
However, the chemical peculiarities
of these stars distort their colour indices, so that their effective temperatures
suffer from an uncertainty which also propagates to other stellar parameters
like e.g. the surface gravity (North \ \cite{n98}). Therefore, it is most
important to determine
the stellar parameters of magnetic stars in the most fundamental way, especially
as the extent to which their evolution within the main sequence is identical
to that of their non-magnetic siblings is not known.

The most suitable systems for the accurate assessment of  fundamental stellar
parameters are double-lined eclipsing binaries. While  
a few double-lined spectroscopic binary systems (SB2)
containing an Ap star of mass  below 3\,M$_\odot$ are currently known,
the rate of close binaries is much smaller among magnetic Bp stars
(Gerbaldi et~al.\ \cite{g85}; Carrier et al.\ \cite{cnub02}; North \& Debernardi \cite{nd04})
and only two double-lined eclipsing binaries with a Bp component,
namely AO\,Vel (= \object{HD\,68826}) and HD\,123335, are known to date
(e.g., Clausen, Gim\'enez \& van Houten \cite{c95}, from here on CGvH;
Hensberge et~al.\ \cite{he04}).
In this paper, we perform a spectroscopic analysis of
the eclipsing SB2 system AO\,Vel consisting of two late B-type stars
with a BpSi primary component (Bidelman \& MacConnell \cite{bi73}),
i.e.\ of a type generally displaying rather large spectroscopic variations,
so that one can infer that there are significant surface inhomogeneities. 
Numerous four-colour photometric observations have been obtained from 1974 to 
1991 with the Danish 50\,cm telescope at La Silla (CGvH).
Times of mid-eclipse for a total of 16 primary and 10 secondary minima have been determined
from all available photoelectric observations and photometric elements and apsidal motion
parameters have been determined from complete $uvby$ light curves.
This study revealed the presence of a third component in a wide 
orbit with an orbital period of 25.6\,yr. The orbital period of the eclipsing system
appeared to be 1.58~days and the orbit was found only slightly eccentric, with a
quite fast apsidal motion (U=56.8\,yr).
Very recently, Wolf \& Zejda (\cite{wz05}) published seven additional
times of minima obtained from the ASAS-3 database (Pojmanski \cite{asas}) and new
photoelectric $UBV$ observations. As a result, they have improved the apsidal
motion parameters and the orbit of the third body, for which they found a period
of 33\,yr.

In spite of the considerable amount of observational data collected for decades, 
absolute dimensions of the components remained unknown since
no accurate radial velocity observations had ever been performed for this system.

\section{Observations and spectroscopic analysis}

\subsection{Observations}

In March 2004 we obtained five spectra of AO\,Vel with a typical
S/N of about 100--150 over three nights with the echelle spectrograph FEROS at
the 2.2\,m  telescope at La Silla, to obtain radial velocities and, ultimately,
the individual  masses of the components. Their wavelength range is 
$3530 - 9220$\,\AA{}, and the nominal resolving power is 48\,000. 
To our surprise, we found that the spectral line profile in the FEROS spectra
exhibit a variable and complex structure indicating four spectral companions at
some phases. In order to confirm the identification of the
four companions and to derive their respective temporal behavior, 
we obtained six additional spectra in January 2005
with the 2.1\,m telescope of the CASLEO  and the echelle 
spectrograph REOSC. These spectra cover the range
3600--6000\,\AA{} at a resolution of 13\,000 and a typical S/N of about 90.
A careful inspection of the spectral line morphology, 
along with the photometric ephemeris published by CGvH, allowed us to 
identify in the spectral line profiles four stellar components
forming two short-period double-lined spectroscopic binaries.
In the following, we identify these stars as A, B, C and D.
The two more massive stars form the eclipsing binary (A+B) with an orbital
period of 1.58~days, the primary component being the BpSi star.
The other spectroscopic binary (C+D) has a period of 4.15~days, and its 
components have a similar spectral type B9--A0.
%% we have to check whether they are A0 and A1 or both AO
% According to Schmidt-Kaler
% T(B9V) = 10500  T(B9III) = 11000
% T(A0V) = 9520   T(A0III) = 10100

\subsection{Spectra separation and RV measurement}\label{sec.rv}

In order to obtain separate spectra for all four components of
the system and to measure their radial velocities (RVs), we applied
the iterative method described by Gonz\'alez \& Levato (\cite{gl05}), 
which was adapted here for multiple systems.
This algorithm  computes the spectra of the individual components
and the RVs iteratively.
In each step the computed spectra are used to remove
the spectral features of all but one component
from the observed spectra. The resulting single-lined 
spectra are used to measure the RV of that component and to compute 
its spectrum by combining them appropriately.
\begin{table*}
\caption{Radial velocities for all four companions.}
\begin{center}
\begin{tabular}{crrrrrrrr} \hline\hline
MJD   &
\multicolumn{1}{c}{$RV_A$}&
\multicolumn{1}{c}{$\epsilon_A$}&
\multicolumn{1}{c}{$RV_B$} &
\multicolumn{1}{c}{$\epsilon_B$}&
\multicolumn{1}{c}{$RV_C$} &
\multicolumn{1}{c}{$\epsilon_C$}&
\multicolumn{1}{c}{$RV_D$} &
\multicolumn{1}{c}{$\epsilon_D$}  \\
           & \kms{}  &  \kms{} & \kms{} & \kms{} & \kms{} & \kms{} & \kms{}  & \kms{}\\ \hline
53070.0187 & -19 & 35 &   -7 & 20 &  -7.3 & 15.0 &  59.8 &  1.0   \\
53071.0875 &-137 &  7 &  150 &  6 & 115.9 &  2.3 & -81.9 &  1.1   \\
53071.2551 &-167 &  7 &  196 &  6 & 115.3 &  2.3 & -83.2 &  1.0   \\
53072.0387 & 174 &  7 & -164 &  6 &  54.2 &  2.2 & -24.3 &  1.1   \\
53072.2683 &  76 &  7 &  -71 &  7 &  34.1 &  2.1 &   9.9 &  1.0   \\
53373.1513 & 143 & 15 & -157 &  6 &  15.2 &  4.3 &  31.2 &  2.8   \\
53373.3296 & 108 &  8 &  -89 &  6 &  51.2 &  3.7 &   2.3 &  2.9   \\
53374.3348 &  63 &  9 &  -38 &  5 & 116.5 &  2.9 & -83.5 &  3.6   \\
53374.2101 & -16 & 11 &   76 &  6 & 120.0 &  3.9 & -80.9 &  2.2   \\
53375.1239 & -42 & 15 &   47 &  6 &  49.3 &  4.1 &   1.8 &  3.3   \\
53375.3616 &-151 & 18 &  180 & 10 &  11.5 &  2.0 &  29.9 &  5.9   \\ \hline
\end{tabular}
\end{center}
\label{rv2}
\end{table*}
\begin{figure*}
%\begin{figure*}[htp!]
\resizebox{\hsize}{!}{
\begin{minipage}[t!]{8cm}
\includegraphics[bb=140 235 500 552,width=5cm,height=11cm]{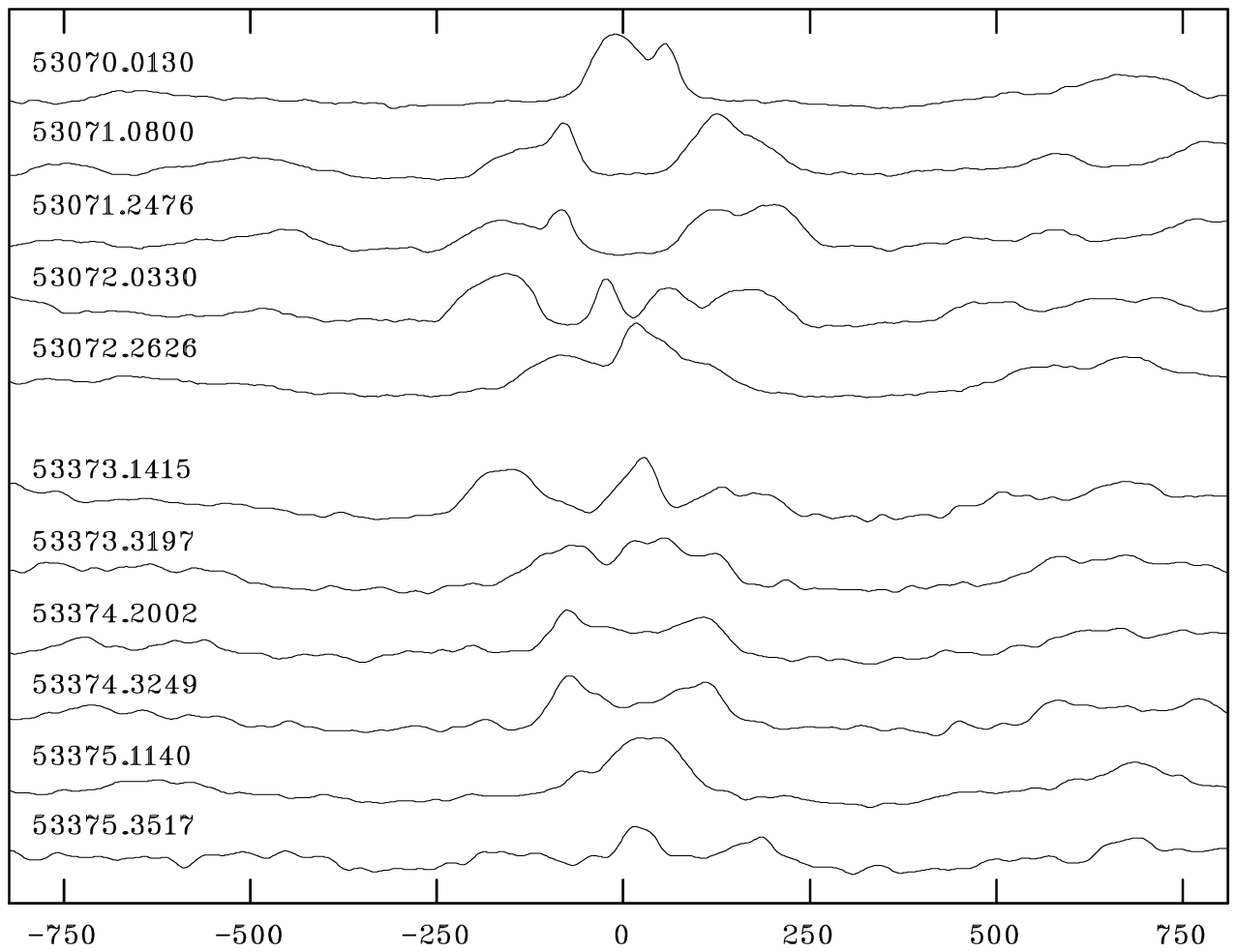}
\end{minipage}
\begin{minipage}[t!]{8cm}
\includegraphics[bb=165 310 630 525,width=10cm,height=4.5cm]{4103fi1b.eps}
\includegraphics[bb=165 330 630 555,width=10cm,height=4.5cm]{4103fi1c.eps}
\end{minipage}}
\caption{Cross-correlation function (left) and  RV curves (right) for
FEROS (top) and CASLEO (bottom) spectra. Filled triangles, open triangles, 
filled circles and open circles represent stars A, B, C and D, respectively.   }
\label{curves}
\end{figure*}
Starting values for RVs were measured in the FEROS spectra by cross-correlation
against spectra of slowly rotating stars of similar spectral type,
HD\,175640 and HD\,196426, which had been observed with FEROS in May 2000.
The RV of these templates were determined by measuring individual spectral lines.
>From these preliminary RVs we calculated the spectra of the four individual
components using our high resolution FEROS spectra.
For these calculations we considered only four of our five FEROS spectra,
since in the first one, taken at MJD\,53070.013, three components appear
strongly blended, making the determinations of RVs less reliable.
The CASLEO spectra also were not considered because of their lower resolution.
During the computation of the spectra of the individual components,
the RVs in those four FEROS spectra were redetermined.

Once the spectra of the four stellar components were obtained, they
were used to compute the RVs from all our 11 observed spectra.
Fig.\,\ref{curves} shows the  RVs of the four stars
as a function of time; the  left panel shows the multiple morphology of the
cross-correlation function.

Table~\ref{rv2} presents the measured RVs for all four stellar components.
Early observations (before MJD\,53373) are those of FEROS.
The assigned RV errors are formal ones calculated within the cross-correlation
analysis. The quality of the RV measurements is also influenced by the
uncertainty in the computed spectra, which are used to remove the lines of the remaining components
from the observed spectra.
In order to evaluate this contribution to the RVs errors,
we calculated the RVs using different subsamples of observed
spectra and in most cases we found that RV differences are comparable with their formal errors.
However, in the spectrum taken on MJD\,53070.0187 the RVs of components A, B and C are
not well defined. The RV computations converge poorly and depend
strongly on the initial velocities. In this case we assigned larger errors so
that all possible solutions are included.

\subsection{Spectral analysis}\label{sec.spec}

%The calculations described in Sec.\,\ref{sec.rv} provided the individual spectra
%of all  four stellar components. However, these spectra are naturally normalized 
%with respect to the combined continuum of the four stars.
%Thus, to recover the correct normalization of the spectrum of each companion, they
To normalize the spectra of individual stellar components we scaled the combined continuum by their
relative luminosity.
We used the relative fluxes in $y$ and $b$ photometric bands 
provided by CGvH for 
the star A and the star B, and the combined flux of C+D. Assuming that the fluxes of stars C and D
are equal, we obtain relative contributions to the observed continuum of
0.40, 0.28, 0.16, and 0.16 for components A, B, C and D, respectively.
The uncertainty of these scaling factors is about 5\%, and should be
considered as a possible systematic error in the intensity of spectral lines.

For each component in the system, we calculated synthetic spectra using the SYNTH code 
(Piskunov \cite{pi92}), the VALD database (Kupka et. al \cite{kup99}) and ATLAS9 model 
atmospheres (Kurucz \cite{kur93}). Starting from models with different
atmosphere parameters, 
we compared a variety of syntheses with observed spectra to determine the best fits 
(Table\,\ref{atmopar}).
Because of the rather poor quality of the spectra, especially of the faint components C and D, 
we mainly took into account the region 4400--4600 \AA \ which includes the
\ion{He}{i} 4471 and \ion{Mg}{ii} 4481 lines, 
as well as the 5000--5100 \AA \,and 6300--6400 \AA \ regions with some prominent Si- and Fe-lines. 
The estimated errors in $T_{\rm eff}$ are of the order of $\pm$ 250\,K for component B and 
$\pm$ 500\,K for components C and D. 
The best choice of microturbulence velocity, $v_{\rm mic}$, was found to be 0 km\,s$^{-1}$ 
for A and B and 2 km\,s$^{-1}$ for C and D.

\begin{figure*}
\resizebox{\hsize}{!}{\includegraphics[bb=80 -70 570 830,angle=-90]{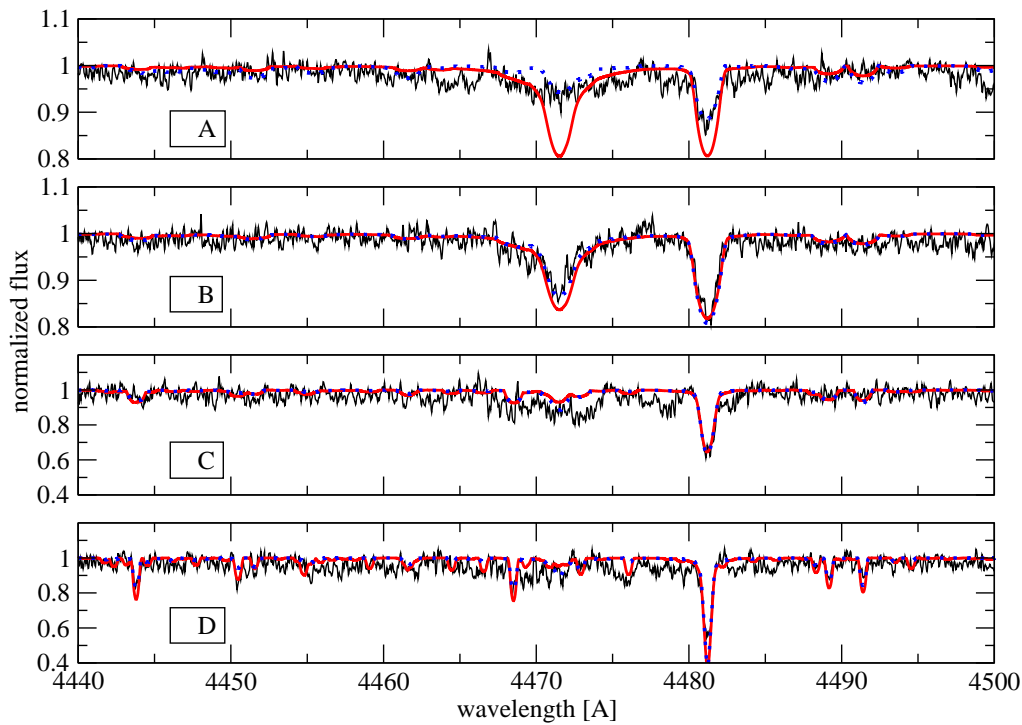}}
\caption{Observed spectra of components A, B, C, and D (thin lines) compared with
synthetic spectra (solid and dotted lines). 
Atmospheric parameters for modeled spectra are as following.
A) Solid line: synthetic spectrum with parameters from stellar models ($T_{\rm eff}$=13700 K, 
${\rm log} g$=4.3) and solar composition, 
dotted line: same model but with Mg and He abundances decreased by 1.45 and 1.04 dex;
B) Solid line: synthetic spectrum with parameters from stellar models 
($T_{\rm eff}$=13200 K, ${\rm log} g$=4.3), 
dotted line: best fit for Mg/He ratio and metal lines ($T_{\rm eff}$=12500 K, ${\rm log} g$=4.3);
C) Solid line: $T_{\rm eff,min}$=10000 K, ${\rm log} g$=4.1,
dotted line: $T_{\rm eff,max}$=11000 K, ${\rm log} g$=4.1;
D) Solid line: $T_{\rm eff,min}$=9500 K, ${\rm log} g$=4.2,
dotted line: $T_{\rm eff,max}$=10500 K, ${\rm log} g$=4.2.\label{synthABCD}}
\end{figure*}

Since the deeper eclipse in the primary system, A+B, corresponds to the occultation of star A,
this star is not only the larger and more massive one,
but also the hotter one. However, \ion{He}{i} lines are weaker than in star B, while \ion{Si}{ii} lines 
are very strong.  
Since the component A is a chemically peculiar star, it is not possible to
derive its temperature
by simply comparing observations with a variety of solar composition models of different
$T_{\rm eff}$ and $\log g$ values. A detailed multi-element abundance analysis would be
needed to determine accurate atmospheric parameters. 
However, such an analysis is not possible yet
because of the unavailability of high-resolution and high signal-to noise spectra. 
For this reason we adopted for the component A the effective temperature predicted by
stellar models for a normal star of the same mass and radius.
Using the Geneva stellar evolution models (Schaller et al. 1992) and the mass and radius calculated
from the light and RV curves (see next section) we found for the A component $T_{\rm eff}$=13700.
In Fig.\,\ref{synthABCD} we show the observed spectrum (thin line) together with the 
synthetic spectrum computed with the atmosphere parameters $T_{\rm eff}$=13700, $\log g$=4.3 and 
solar element abundances (solid line). 
In order to reproduce the \ion{He}{i}  and \ion{Mg}{ii} features in the 4470--4480 \AA \, 
region by the synthesis using the same atmosphere parameters 
the abundances of He and Mg had to be decreased by 1.45 and 1.04 dex, 
respectively (dotted line). From the \ion{Si}{ii}  lines in the spectrum of star A we derived 
a Si-overabundance of $\approx$ 0.5 dex relative to the solar composition model, which is in good
agreement with the classification of this star as BpSi by Bidelman \& MacConnell (\cite{bi73}).
Star B is a B8\,V normal star (Fig.\,\ref{synthABCD}). The observed spectrum can be very 
well fitted by a synthetic spectrum assuming solar composition, 
$T_{\rm eff}=12500\pm 250$\,K and $\log g =4.3 \pm 0.1$.
			                                                                        
The spectra of stars belonging to the spectroscopic binary C+D are 
similar to each other and correspond approximately to spectral type B9--A0. 
The spectral analysis of the system C+D using the SYNTH
code indicates $T_{\rm eff}=10500\pm 500$\,K for the star C (Fig.\,\ref{synthABCD}), 
which is the more massive companion, 
and $10000\pm 500$\,K for the less massive companion D (Fig.\,\ref{synthABCD}). 
The \ion{He}{i} line in the spectra of C and D cannot be considered 
as a reliable temperature indicator since the continuum determination 
in this region is rather uncertain. The uncertainty of our temperature estimates, which are
based only on a few well resolved, unblended Fe lines, is rather high.
The majority of Mg and Fe lines in the spectrum of star C can be modelled rather well assuming a 
purely solar composition, $T_{\rm eff}$=10500 K and $\log g=4.1\pm 0.2$. 
While most metal lines in the spectrum of star D can be well reproduced by the synthesis assuming
solar composition, $T_{\rm eff}$=10000 K and $\log g=4.2\pm 0.2$, the observed \ion{Mg}{ii} line 
at 4481 \AA\, indicates a lower abundance ($\approx$ -0.4 dex) than predicted by the model.   
%%%Here is a table of parameters used for spectrum synthesis:
\begin{table}
\caption{Atmospheric parameters derived from comparison with synthetic spectra. \label{atmopar} }
\begin{tabular}{l  c c c} \hline \hline
Star   &  $T_{\rm eff}$& $\log g$& $v\,\sin i$   \\
       &  [K]                      &                 & [km\,s$^{-1}$]\\ \hline
 A          & 13700$^{\mathrm{a}}$                    & 4.3             & 65            \\
 B          & 12500  $\pm$ 250\,K      & 4.3 $\pm$ 0.1    & 73            \\
 C          & 10500  $\pm$ 500\,K      & 4.1 $\pm$ 0.2    & 40            \\
 D          & 10000  $\pm$ 500\,K      & 4.2 $\pm$ 0.2    & 18            \\ \hline
\end{tabular} \\ 
\begin{list}{}{}
\item[$^{\mathrm{a}}$] Adopted from stellar models according to the observed mass and radius.
\end{list}
\end{table}

For a more careful spectral analysis a larger number of observed spectra at different orbital phases
is needed, not only to improve the signal-to-noise ratio but also to assure a better disentangling
of spectral components.

\section{Orbital analysis}

\subsection{Spectroscopic orbits}

Spectroscopic orbits for the two binaries A+B and C+D
were calculated using the 11 RV measurements of each star.
Weights were given to the RV measurements according to their uncertainties. 

For the system C+D we used the least squares method to determine the following 
parameters: P$_{CD}$ (orbital period), 
$K_C$, $K_D$ (RV amplitudes), $V\!\gamma_{C\!D}$ (center-of-mass RV), $e_{C\!D}$ (eccentricity), 
$\omega_{C\!D}$ (argument of the periastron), and $T_{C\!D}$ (time of periastron passage).
The RMS of the residuals is 4.9 \kms{} for star C, while for star D it is 1.0 \kms{} for La Silla observations
 and 4.1 \kms{} for CASLEO spectra.
These values are consistent with the measurement errors.

In the case of the  system A+B, we fitted the RV curves
using the Wilson \& Devinney program
(Wilson \& Devinney \cite{wd71}, Wilson \cite{wil90}) 
in order to take into account the proximity effects which are
significant in such a short-period binary.
We lowered the weight of observations taken during the eclipses by
a factor of 2, since the effect of partial eclipse on the RVs (Rossiter effect, 
Rossiter 1924)
% This effect was incorporated in the light&RV curves modeling by
% Wilson (1979ApJ...234.1054)
has not been specifically modelled during our calculation of the component
spectra and RV measurements. 

A detailed study of the spectral line profile variability during the eclipses
would be especially interesting to obtain information about surface
inhomogeneities of chemical elements in the BpSi primary.
However, a larger number of high-resolution spectra are needed 
for such a study, which is thus deferred to a future work.

Reasonable values for the atmospheric parameters 
required by the Wilson \& Devinney program were assigned according to the spectral types.
We adopted the relative radii and orbital inclination 
calculated by CGvH from the analysis of
the photoelectric {\em uvby} light curves of Gr\o nbech (\cite{g87}).
We let the epoch T$_{AB}$ as a free parameter since the photometric ephemeris 
calculated by CGvH were not compatible with our RV curves.
As described in the next section, 
this and other inconsistencies led us to reanalyze the photometric
ephemeris and to propose parameters of the long-period orbit that
are quite different from those found by previous works.
The orbital parameters resulting from the RV curve analysis
are listed in Table\,\ref{sppar}.

\begin{table}
%\begin{table}[htp!]
\caption{Spectroscopic orbits of binary systems C+D and A+B.}
\begin{tabular}{lcrcl} \hline\hline
Parameter   &  Units & \multicolumn{3}{c}{Value}    \\ \hline

$P_{C\!D}$              & days &   4.15008 &$\pm$&0.00016   \\
$T_{C\!D}({\rm periastron})$& MJD&  53074.059 &$\pm$&0.060   \\
$V\!\gamma_{C\!D}$      &\kms{} &     21.2  &$\pm$&0.7    \\
$K_C$                   &\kms{} &    98.3   &$\pm$&2.1   \\
$K_D$                   &\kms{} &    107.4  &$\pm$&1.3      \\
$\omega_{C\!D}$         &deg  &    60     &$\pm$&   6    \\
$e_{C\!D}$              &     &    0.047  &$\pm$&0.015   \\
$a_{C\!D}\sin i_{C\!D}$              &\Rsun &   16.84  &$\pm$& 0.23 \\
$q_{C\!D}$              &      &   0.916  &$\pm$& 0.024 \\
$M_C \sin^3 i_{C\!D}$   &\Msun{} &   1.94   &$\pm$&0.07     \\ 
$M_D \sin^3 i_{C\!D}$   &\Msun{} &   1.77   &$\pm$&0.08      \\ \hline
%\multicolumn{5}{c}{System A+B} \\ \hline
$P_{A\!B}$              & days  & 1.584584$^{\mathrm{a}}$& &   \\
$T_{A\!B}({\rm periastron})$ &MJD & 53069.8821&$\pm$&0.0055     \\
$T_{A\!B}({\rm Min\,I})$ &MJD & 53070.8440&$\pm$&0.0055     \\
$V\!\gamma_{A\!B}$ &\kms{} &      8.9 &$\pm$& 2.3      \\
$q_{A\!B}$ &		& 0.931 & $\pm$& 0.033 \\
%K$_1$ (\kms{}) &    170.7   & $\pm$&3.8         \\
%K$_2$ (\kms{}) &    182.4   & $\pm$&3.3         \\
$\omega_{A\!B}$  &deg& 236.5$^{\mathrm{a}}$ &&    \\
$\dot{\omega}$   & deg yr$^{-1}$& 0.0174$^{\mathrm{a}}$  & &  \\ 
$e_{A\!B}$  &        & 0.0741$^{\mathrm{a}}$ &&  \\
$i_{A\!B}$  & deg   & 88.5$^{\mathrm{b}}$  & & \\
M$_A$ &\Msun{}       &   3.63  &$\pm$& 0.18      \\
M$_B$ &\Msun{}       &   3.38  &$\pm$& 0.18  \\ 
a$_{A\!B}$ &\Rsun  &  10.94  &$\pm$& 0.19 \\ 
R$_A$ &\Rsun       &   2.34$^{\mathrm{c}}$  &$\pm$& 0.08 \\
R$_B$ &\Rsun       &   2.11$^{\mathrm{c}}$  &$\pm$& 0.08 \\  \hline

\end{tabular}
\begin{list}{}{}
\item[$^{\mathrm{a}}$] Calculated from photometric ephemeris analysis
 (Sec.\,\ref{ephem}). $P_{A\!B}$ and $\omega_{A\!B}$ are calculated for MJD\,53070.
\item[$^{\mathrm{b}}$] Adopted from CGvH.
\item[$^{\mathrm{c}}$] Computed using the photometric relative radii 
published by CGvH.
\end{list}
\label{sppar}
\end{table}

\subsection{The long-period orbit between the two binaries}
\label{ephem}

To better constrain the fundamental parameters of the AO\,Vel system, we did
an in-depth revision of all published data and analyzed all 
available times of minima of the eclipsing binary.
In the study of the eclipsing system A+B,
CGvH fitted the epochs of 26 eclipses measured throughout 26 years
using a period of 1.584653 $\pm$ 0.000005 days and considering the
combined effect of the apsidal motion of the orbit at a rate of
0.0275$^\circ$/cycle and the light time effect due to the orbit 
with a third body (our binary system C+D).
However, the time of minimum predicted for the epoch of our observations 
differs from our spectroscopic ephemeris by 0.060 $\pm$ 0.026 days.
In addition, according to the ``third body'' orbit found by CGvH
for the time of our FEROS observations 
the barycentric RV of the binary A+B should be higher than
that of the binary C+D for La Silla and CASLEO observations,
which is absolutely incompatible with the observed RVs since,
according to our measurements,
$V\!\gamma_{A\!B} - V\!\gamma_{C\!D}  = - 12.3 \pm 2.4 $ \kms{}.
Finally, the  third body orbit of CGvH
is unable to match the 38 old photographic observations taken by
Oosterhoff \& van Houten (\cite{OvH49}) between the years 1902 and 1935.
CGvH have admitted this disagreement in their paper and
speculated that might be a consequence of
some kind of longer time scale period variation.

On the other hand, the parameters recently published by Wolf \& Zejda (2005)
for the wide orbit are very different from those of CGvH.
The ephemeris calculated by  Wolf \& Zejda (2005) 
agree with our observations within 0.003 days.
They found an eccentric orbit ($e=0.21$) with a period of 33.3 yr, which
might account for the observed RV difference between
the two binaries. However, it is also in disagreement with the
old photographic data.

For these reasons, we decided to recompute the apsidal motion
and the third body orbit using all the available times of minimum 
along with our spectroscopic observations. 
In view of the very different quality of the various photometric sources,
we have given weight 1 to the photoelectric minima (26 from CGvH and one
from Wolf \& Zejda 2005), weight 0.1 to the ASAS-3 data, and weight 0.01
to the much less precise  photographic data. 
We  have also included the time of minimum
predicted by our spectroscopic orbit as a single point 
with unit weight.

The key difference with respect to the CGvH
calculations is that we do not assume the wide
orbit to be circular. On the other hand, we have included the
photographic data, which were not considered by Wolf \& Zejda (2005).
Consequently, we were able to
find a solution that is compatible with both our
spectroscopic orbit and the old photographic observations.
Figure~6 shows our fit of the times of minima  variation due
to the 3rd body (cf.\ with Fig.~4 in CGvH and Fig.\,6 in Wolf \& Zejda 2005).
The RMS of the residuals was 4$\times$10$^{-4}$ days for the
photoelectric minima, 7$\times$10$^{-3}$ days for the ASAS-3 photometry 
and 2.6$\times$10$^{-2}$ days for the photographic data.

In addition to the determination of the parameters of the orbit AB+CD, the
ephemeris fitting provided simultaneously the orbital period
of the eclipsing binary and its apsidal motion rate.
Table\,\ref{tabcd} shows the results obtained.
The parameters we found for the apsidal motion are 
indistinguishable from  those found by CGvH and Wolf \& Zejda (2005).
\begin{figure}
\resizebox{\hsize}{!}{\includegraphics{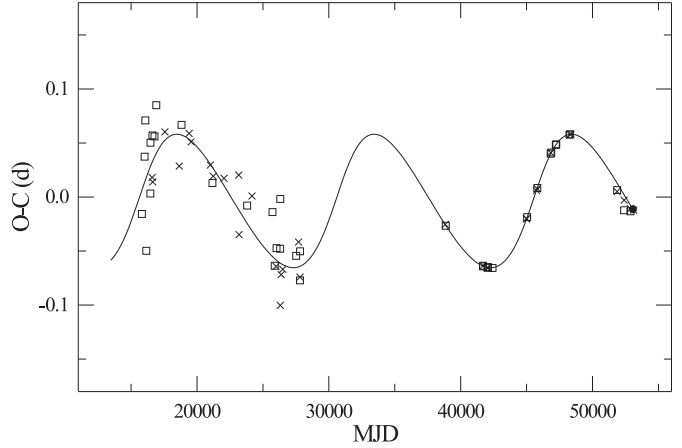}}
\caption{Times of minimum shifts caused by the third body
orbit. Symbols are the differences between the observed times
of minima  and the  times of minima
calculated using mean ephemeris, after correction for
apsidal motion (primary eclipses are marked by squares, secondary eclipses
are marked by crosses and
the filled circle shows Min I calculated by fitting our RV curves).}
\end{figure}
However, the parameters defining the orbit AB+CD are essentially different
from those found by CGvH (P$_o$ = 25.6 $\pm$ 2.5 yr; 
$a_{o,1} \sin i_o$ =  0.041  $\pm$ 0.006 days) or by Wolf \& Zejda (2005)
(P$_o$ = 33.3 $\pm$ 0.5 yr;
$a_{o,1} \sin i_o$ =  0.052  $\pm$ 0.004 days),
and also the binary period $P_{A\!B}$ differs significantly.

\begin{table}
\caption{Orbital parameters derived from the analysis of times of
minima. \label{tabcd} }
\begin{tabular}{lcrcl} \hline \hline
Parameter   &  Units & \multicolumn{3}{c}{Value}    \\ \hline
\multicolumn{5}{c}{Orbit  A+B} \\ \hline
$P_{A\!B}$      & days    &    1.5846154& $\pm$&0.000002   \\
$T_{A\!B}({\rm Min I})$& MJD & 45043.6608&$\pm$&0.0004     \\
$P_{\rm aps.\,motion}$ &yr  &   54.72& $\pm$ &0.45   \\
$T_{\omega=0}$ &MJD  &  39490& $\pm$& 90 \\ 
$e_{A\!B}$ & & 0.0741 & $\pm$& 0.0004 \\ \hline
\multicolumn{5}{c}{Orbit AB + CD} \\ \hline
P$_o$ &yr     & 41.0& $\pm$& 0.2      \\
$T_o({\rm periastron})$ &    MJD &45839& $\pm$ &11      \\
$T_o({\rm conjunction})$ &    MJD& 47803 &$\pm$& 120     \\
$a_{o,1} \sin i_o$ &days  & 0.0642& $\pm$ &0.0004     \\
$a_{o,1} \sin i_o$ &AU    &11.15 &$\pm$& 0.07     \\
$e_o$  &  & 0.291& $\pm$& 0.005       \\
$\omega_o$& deg  &   11.8 &$\pm$& 1.4   \\ \hline
\end{tabular} \\
\end{table}
 
With the new parameters for the wide orbit the eclipsing binary is 
at present moving
toward the observer, as expected according to the observed RVs.
The predicted RV  for the eclipsing binary is $-$6.0
\kms{} with respect to the center-of-mass of the quadruple system,
which is consistent with the observed velocities (see Sec.\,\ref{sec.par}).
However, more  accurate RVs (errors of about 0.3--0.7 \kms{} in $V\!\gamma_{A\!B}-V\!\gamma_{C\!D}$)
would be required for the  mass-ratio
$M_{C+D}/M_{A+B}$ to be calculated with reasonable precision (5--10 \%).

\section{Physical parameters}

\subsection{Stellar parameters and evolutionary state\label{sec.par}}

\begin{figure}
\resizebox{\hsize}{!}{\includegraphics{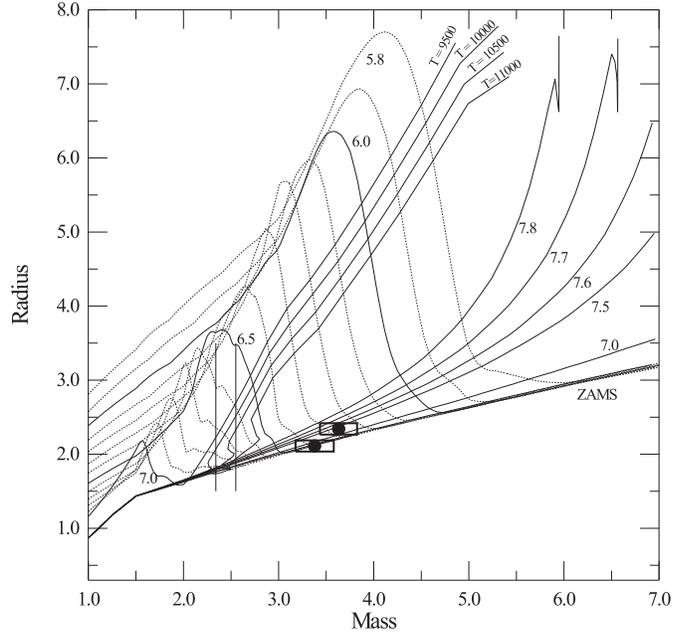}}
\caption{ Mass-radius diagram. Filled circles with error boxes are stars A and B
belonging to the eclipsing system.
Minimum masses for stars C and D are marked with vertical bars.
Main sequence (Schaller et~al.\ \cite{Sc92}, continuous lines) and
pre-main sequence (Bernasconi \cite{Be96}, dotted lines) isochrones are plotted
and labeled with log(age). Thick lines are isotherms interpolated for
pre-main-sequence models.
}
\label{mr}
\end{figure}

For the eclipsing pair we have derived absolute masses and dimensions 
using the photometric elements (inclination and relative radii) published
by CGvH along with our spectroscopic data.
In the mass-radius diagram shown in Fig.\,\ref{mr} both A and B are located near the ZAMS.
%being TAMS radius for these masses  about 5 \Rsun.
For comparison, isochrones of the Geneva stellar models
(Schaller et~al.\ \cite{Sc92}) for solar composition 
are also plotted in Fig.\,\ref{mr}.
The mass and radius of star A indicates that the age of the system is less than
about 5 $\times 10^7$ yr, being compatible with the ZAMS.
This corresponds to less than 25 \% of the main sequence lifetime for star A.

Since the orbital inclination for the binary C+D is unknown,
the individual masses of these stars cannot be derived from
the RV curves.
However, some information can be obtained from the outer orbit
between the two systems and the mass of the binary A+B.
In the following we use subindex ``$o$'' for the parameters of the orbit binding
both binary systems. In particular,
M$_o$ is the total mass of the quadruple system and q$_o$ = M$_{C+D}$/M$_{A+B}$.

>From the ephemeris analysis of the eclipsing binary we have found
the parameter $a_{o,1} \sin i_o$, which is related to
the semiaxis of the relative orbit ($a_o$) through:
$$
a_o \sin i_o = a_{o,1} \sin i_o \cdot (1+q_o^{-1}). 
$$
Combining this expression with the third Kepler equation
$$
M_{o} = M_{A+B}\cdot (1+q_o) = a_o^3 \cdot P_o^{-2},
$$
we obtain:
$$
 \frac{(1+q_o)^2}{q_o^3} = (a_{o,1} \sin i_o)^{-3}\cdot P_o^2 \cdot M_{A+B} \cdot \sin^3 i_o,
$$
where $(a_{o,1} \sin i_o)$ is given in AU and $P_o$ in years.
For a given value of $i_o$ this equation provides
$q_o$, and hence the absolute masses of the system C+D: \\
$M_D = q_o \cdot M_{A+B} / (q_{C\!D}+1)$, $M_C = q_{C\!D}\cdot M_D$.
>From the condition $i_o \leq  90^\circ$ a lower limit for the masses of
components C and D can be established. 
The resulting minimum masses are 2.55$\pm$0.09\,\Msun{} and 2.34$\pm$0.10\,\Msun{}. 
%which are higher than those expected from their spectral types determined from their spectra.
%In other words, these stars have lower temperatures than main-sequence
%stars of the same mass. 
%This fact would suggest that the C and D components are pre-main sequence stars.

Using these minimum masses for the stars C and D and their spectroscopic orbit, 
we obtain the orbital inclination of the non-eclipsing binary to be
$i_{C\!D} \leq 66.0 \pm 2.4$ deg.
If the mass of C+D is close to their lower limit, then
$q_o \approx 0.70$, and the expected RV difference between
the two binaries is
$V\!\gamma_{A\!B} - V\!\gamma_{C\!D} = -14.5 \pm 0.6$ \kms{}, in agreement with
the observed value ($-12.3 \pm 2.4$ \kms{}). 

\begin{figure}
%\begin{figure}[htp!]
\resizebox{\hsize}{!}{\includegraphics{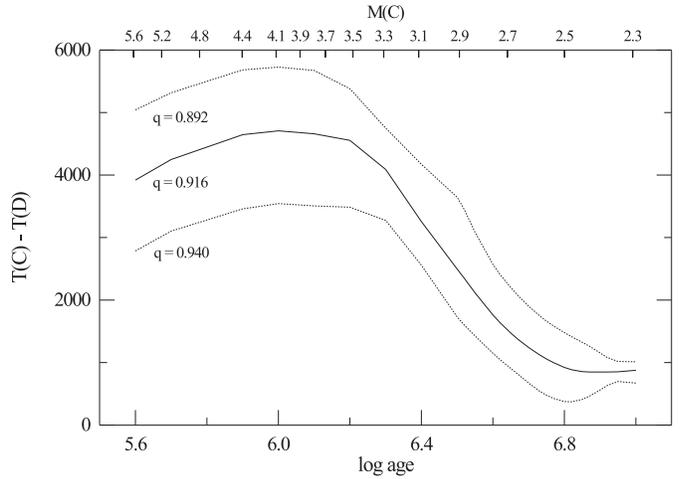}}
\caption{Expected temperature difference of stars C and D as
a function of age, according to pre-main sequence models
of Bernasconi (\cite{Be96}). 
}
\label{Tdif}
\end{figure}

The parameters of the wide orbit restrict the position of
stars C and D in the mass-radius diagram to the right of the 
two continuous vertical lines in Fig.\,\ref{mr}.

Assuming the four stars to be coeval, stars C and D would be very close to
the ZAMS, considering the youth of the more massive stars.
Even the possibility of them being pre-main sequence stars must be considered.
In Fig.\,\ref{mr} pre-main sequence isochrones for solar composition
from  Bernasconi (\cite{Be96}) are plotted.
The spectroscopic temperatures and the luminosity difference
between the two binary systems provide useful constraints
on the evolutionary state of the stars C and D.
Three isotherms have been interpolated in the pre-main sequence models
of Bernasconi (\cite{Be96})
at temperatures 9500, 10000, 10500, and 11000 K. These curves define in
Fig.\,\ref{mr}  two overlapping strips corresponding to stellar models with
temperatures close to those inferred from the spectra of components
C and D (9500--10500 K and 10000--11000 K, respectively).
Within these strips, the younger massive models are not compatible
with the observations for various reasons.
First of all, if all four stars are assumed to be coeval,
pre-main sequence ages of the order or below  10$^{6.3}$\,yr should be discarded
because of the small radius of star B.
Second, massive models for stars C and D would be too luminous
with respect to components A and B. 
%%%%%%%%%%%%%%%%%%%%%%
More precisely, models with log(age)$<$6.55 are not compatible with 
the light-ratio derived from the light curve analysis by CGvH.  
%According to the light curve analysis
%by CGvH, the light-ratio CD:AB in the $y$ and $b$ Str\"omgren bands
%is 0.475 $\pm$ 0.022. This implies that, even assigning conservative
%uncertainties to the bolometric corrections estimated from the spectral types,
%it can be asserted that the luminosity-ratio between the two binaries is 
% L$_{C\!D}$/L$_{A\!B} < 0.41$. 
%This condition imposes a lower limit of about 10$^{6.55}$ yr to the age, 
%corresponding to $M_C < 2.8$ \Msun{}.
Finally, in the upper portion of the temperature strips in Fig.~\ref{mr}
two stars with a fixed mass-ratio $M_D/M_C = 0.916$ would have a 
temperature difference $T_C - T_D$ larger than expected. 
This is a consequence of the rapid evolution at
larger masses and holds also for the difference in luminosity or radius.
As illustration, Fig.\,\ref{Tdif} shows the temperature difference $T_C - T_D$
as a function of age.
Each curve corresponds to the temperature difference of two pre-main sequence
stellar models with a given mass ratio (as labeled) and with an average 
temperature $(T_C + T_D)/2 = 10250$ K.
Three mass-ratio values have been considered to take into account the
observational error of $q_{CD}$. 
The similarity of the spectral types of stars C and D favors again the older models.
In conclusion, stars C and D are most probably on the ZAMS or only
slightly younger. The age of the system is  not younger than about $10^{6.6}$\,yr
in the pre-main sequence grid of Bernasconi (\cite{Be96}).

Additional spectroscopic observations would help to determine
the age with higher precision by improving the calculation of stellar parameters
for the eclipsing system and the temperature estimates for the less massive binary.
In particular, if the pre-main sequence nature of stars C and D was
established, then the post-main sequence evolutionary age of the peculiar star A 
would be less than about 4\% of its main-sequence lifetime.

An additional argument in favor of the youth of the AO\,Vel system arises from
the apsidal motion rate.
This rate can be used in combination with the stellar and orbital parameters
to derive the average value of the internal structure constant
$k_2$ of the components. 
The influence of the  system C+D on the apsidal motion of the
binary A+B is negligible due to the long period of the outer orbit.
Our calculations give $\log k_2 = -2.25 \pm 0.05$. This is 
essentially the same value found by CGvH, since this
parameter depends strongly on relative radii but only slightly
on the masses.
Theoretical models for the solar composition
(Claret \& Gim\'enez \cite{cg92}) predict
$\log k_2 = -2.18 \pm 0.02 $ for ZAMS stars of the same mass.

\subsection{Absolute magnitude and distance}

Even though absolute masses and dimensions are not precisely known for
stars C and D, their contribution to the total luminosity of
the quadruple system is rather small, so that
the absolute magnitude of the latter
can be estimated with reasonable precision.
We have derived the absolute visual magnitudes of the components by interpolation in the stellar
model grids according to their masses. 
If all four stars are assumed to be close to the ZAMS 
the integrated absolute magnitude of the quadruple is
$M_v = -0.82 \pm 0.12 $.
This magnitude would be somewhat brighter if the system was more evolved ($M_v =-0.90$
for $\log \tau = 7.5$) or if components $C$ and $D$ were assumed to be still
PMS stars ($M_v = -0.95$ for  $\log \tau_{PMS}$ = 6.6). 
Using the apparent magnitude $V=9.34$, we obtain an 
apparent distance modulus of 10.22 $\pm$ 0.16.

AO\,Vel is located in the region of the Milky Way populated
by several OB associations and sparse open clusters: open cluster Cr 173,
OB association identified by Brandt et~al.\ (\cite{b71}) and later referred to
as Vel OB2, the ``Vela sheet'' (Eggen \cite{e80}, \cite{e83}; Loktin et~al.
\ \cite{lot1}; Loktin \& Beshenov \cite{lot2}) behind which lie several young
associations embedded in a dust cloud (Kaltcheva \& Hilditch 2000).
AO\,Vel might be a member of the Vel OB4 association identified by the latter
authors (and probably related with the Vel OB2 association of
Slawson \& Reed 1988), the mean distance of which is 0.72 kpc.
% In this case the reddening for AO\,Vel should be 0.31 mag.
% I have no good estimation of the reddening for AO\,Vel.
% 
%% It is not clear from the sentence above what S&R have identified.
% i.e. m-M = 10.3-12.1
% In my opinion there is a  serious problem with
% the identification of star associations in the region.
% Kaltcheva \& Hilditch (2000) identify their VelOB2
% association with VelOB2 of Slawson \& Reed (1988), and
% the same for VelOB4. But Slawson \& Reed (who studied
% only faint stars in a small region) did not found
% stars belonging to the Vela Sheet. All their stars have 
% E(B-V)>0.4 (are already inside the cloud).
% So, their first group at 650 pc (they call it OB2)
% should be associated with the OB4 of  Kaltcheva \& Hilditch (2000).
% The small group at 350 pc in Kaltcheva \& Hilditch (2000)
% (called OB2) does correspond to the Vela sheet.
% In conclusion:
% -  AO\,Vel is certainly non-member of
%    the vela sheet (Cr 173, Gamma velorum system, the
%    Brandt et al's association, Kaltcheva's OB2), 
% -  It is not a member of the most embedded groups (OB1, OB3)
% -  It is a possible member of the first OB association
%    embedded in the cloud, which (in my opinion) corresponds
%    to Kaltcheva's OB4 and Slawson's OB2.
%
%
%Note: Eggen (1983) includes AO\,Vel in his study and mention
%that it might be a member of Vel OB1. But unlike the other authors
%he calls OB1 to the closest and OB2 to the more distant.

We note that if the distance to AO\,Vel is of the order of 700 pc, this system might be
resolved astrometrically as a double star.
The maximum separation between the two spectroscopic binaries would be then about 27 AU, 
corresponding to about 0.04\,arcseconds.
This is too small for the Hipparcos mission --- which took
place at the time of conjunction (MJD 47700), not of quadrature --- but
accessible to interferometers like the VLTI. The maximum projected separation
occured on MJD 52740 (April 2003), so that further astrometric observations of
this multiple are strongly encouraged.

\section{Conclusions}

Using our recent spectroscopic observations we discovered that the triple system
AO\,Vel with an eclipsing BpSi primary is in fact a remarkable quadruple system
close to the ZAMS. 
The analysis of our data as well as all available published data revealed completely 
different orbital parameters for this system compared to previous works.
We found that
both spectroscopic binaries move in a wide eccentric orbit ($e=0.29$) with a period
of 41\,yr and this orbit is fully consistent  with the observed difference 
between the center-of-mass radial velocities of the two spectroscopic binaries.
The calculated orbit imposes a lower limit to the total mass of the binary C+D:
M$_C$+M$_D$ $\ge$ 4.9\,M$_\odot$. 
% actually 4.89 \pm 0.18, so it could be expressed as M$_C$+M$_D$ $\ge$ 4.7 
The analysis of the stellar parameters and of the evolutionary state of the
components reveals their extreme youth.
Since no magnetic field measurements have been performed for this system,
we have applied for VLT time to obtain spectra in
circular polarized light with FORS\,1, which is 
equipped with polarization analyzing optics.

We also plan to obtain additional high resolution, high signal-to-noise spectra of this system
at different orbital phases, in order to improve the
determination of the stellar parameters and atmospheric elemental abundances
of all companions. This unique system is especially important to test
evolutionary models of very young stars of intermediate mass.
Detailed observations during primary eclipses will provide strong 
constraints on the distribution of the abundance anomalies on the surface of the
A component, complementing the standard Doppler imaging technique based on
rotation.
  
\begin{acknowledgements}
We thank Claudio Melo for the reduction of one FEROS spectrum.
\end{acknowledgements}

%\newpage

\end{document}